\documentclass[aps,showpacs,showkeys,superscriptaddress,amssymb,amsmath,amsfonts]{revtex4}
\usepackage{epsfig}
\begin{document}

\title{\bf Absolute Branching Ratio Normalization for Rare \\
\boldmath $\pi^+$ and $\mu^+$ Decays in the PIBETA Experiment}

\newcommand*{\uva}{University of Virginia, Department of Physics, Charlottesville, VA 22904, USA}
\affiliation{\uva}

\author{Emil~Frle\v{z}\footnote{Tel: +1--434--924--6786, 
FAX: +1--434--924--4576, e--mail: frlez@virginia. edu\hfill} } 
\affiliation{\uva}
\collaboration{The PIBETA Collaboration}\noaffiliation

\begin{abstract}
We have used the PIBETA detector at the PSI for a precise measurement of 
rare pion and muon weak decays. We have collected a large
statistical sample of (1) $\pi^+\to e^+\nu_e$,
(2) $\pi^+\to \pi^0 e^+\nu_e$,
(3) $\pi^+\to e^+\nu_e\gamma$, 
(4) $\mu^+\to e^+\nu_e\bar{\nu}_\mu$, 
and (5) $\mu^+\to e^+\nu_e\bar{\nu}_\mu\gamma$ decays.
We have evaluated the absolute branching ratios for
these processes by normalizing to the independently
measured number of decaying $\pi^+$'s (or $\mu^+$'s). We discuss
the mutual consistency of the preliminary results.
\end{abstract}
\pacs{12.15.Hh, 13.20.Cz, 14.35.Bv, 14.40.Aq, 14.60.Ef}
\keywords{determination of CKM matrix elements, decays of pions and muons,
properties of pions and muons}
\maketitle

\begin{figure} [h] 
\begin{center}
\includegraphics[scale=0.40,origin=c]{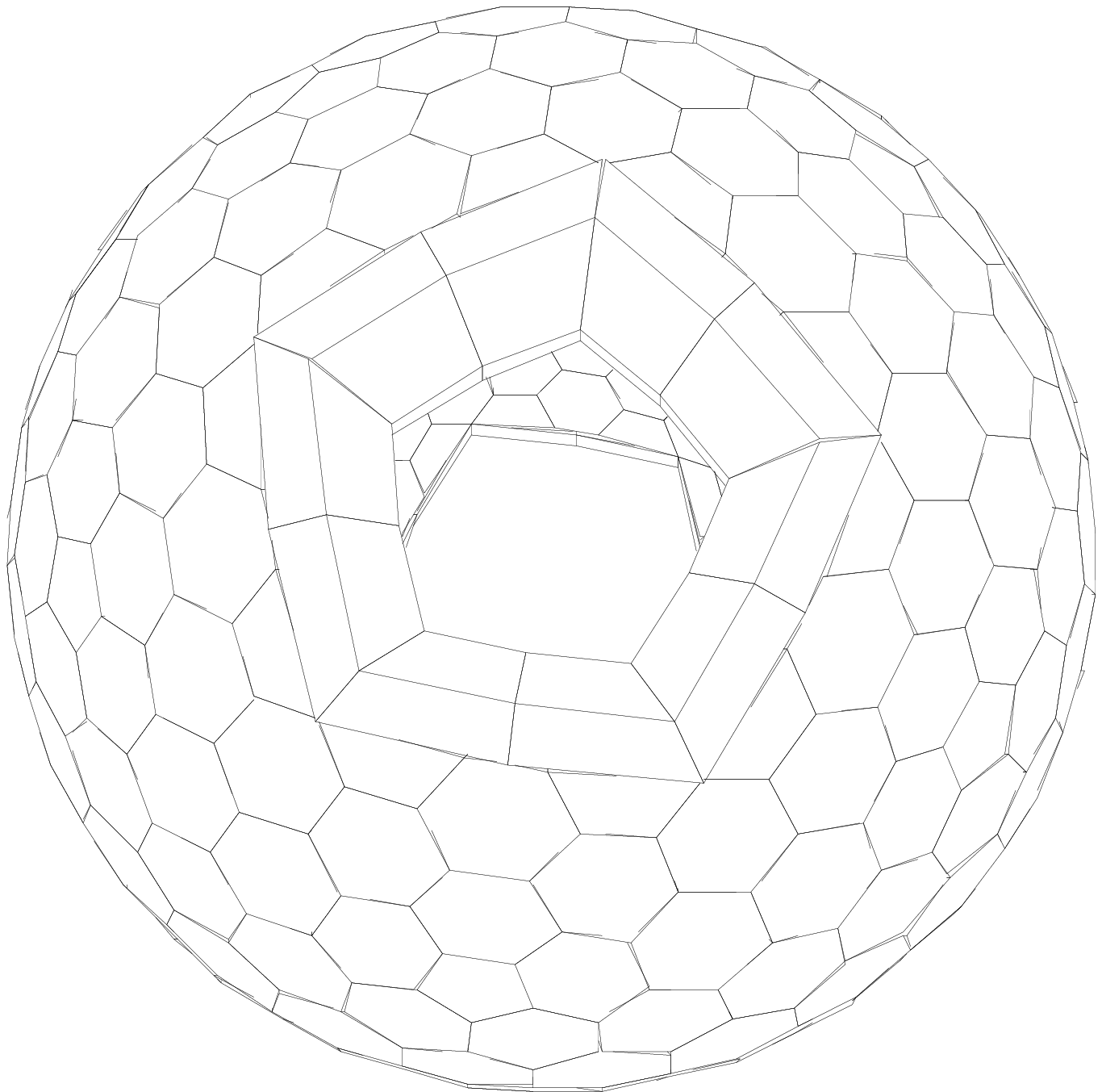}\hspace{1.5cm}
\includegraphics[scale=0.35,origin=c]{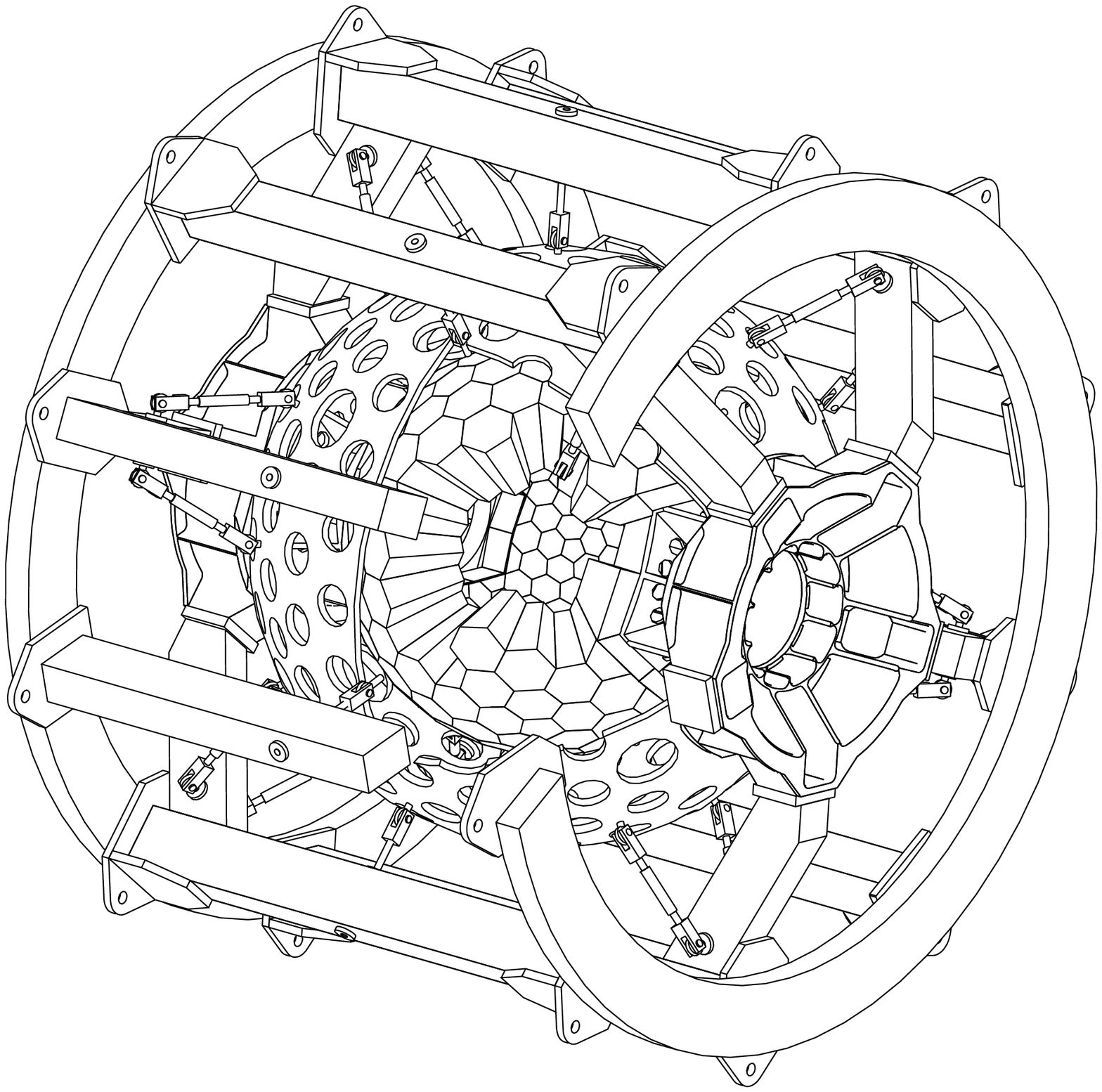}
\end{center}
\caption{The schematic drawing of the {\sc PIBETA} detector. The left 
panel shows the geometry of the 240-module pure CsI shower calorimeter.
The technical drawing of the assembled calorimeter is shown in the
right.}
\label{fig:pb_det}
\end{figure}

The PIBETA Collaboration~\cite{coll} at PSI has performed a series of high 
precision measurements of rare pion and muon decays.
We have used the PIBETA detector~\cite{pb}, a non-magnetic,
segmented, pure CsI spherical calorimeter supplemented with 
a pair of cylindrical multi-wire proportional chambers for
the charged particle tracking and a plastic veto hodoscope
for the particle identification (Fig~\ref{fig:pb_det}).

The primary goal of the experiment has been to determine
the pion beta decay ($\pi^+\to \pi^0 e^+\nu_e$) branching ratio
with $\sim 0.5\,$\% uncertainty, improving the precision
of previous measurements by almost an order of magnitude~\cite{McF85}.
Pion beta decay provides the theoretically most unambiguous 
means to study weak $u$-$d$ quark mixing which directly tests 
quark-lepton universality and can thus constrain certain aspects 
of physics beyond the present Standard Model. 

In the PIBETA experiment a total of $2.2\cdot 10^{13}$ $\pi^+$ beam 
stops were recorded during several running periods spanning three years. 
The beam pions were counted by a tight fourfold coincidence between (1) a forward beam 
counter BC, (2) active degrader AD, (3) active target AT, and 
(4) rf accelerator signal. The non-pionic beam contamination 
determined by the time-of-flight method was small, 
0.4$\,$\% $e^+$'s and 0.2$\,$\% $\mu^+$'s, respectively.

\begin{figure} [h] 
\begin{center}
\includegraphics[scale=0.35,origin=c]{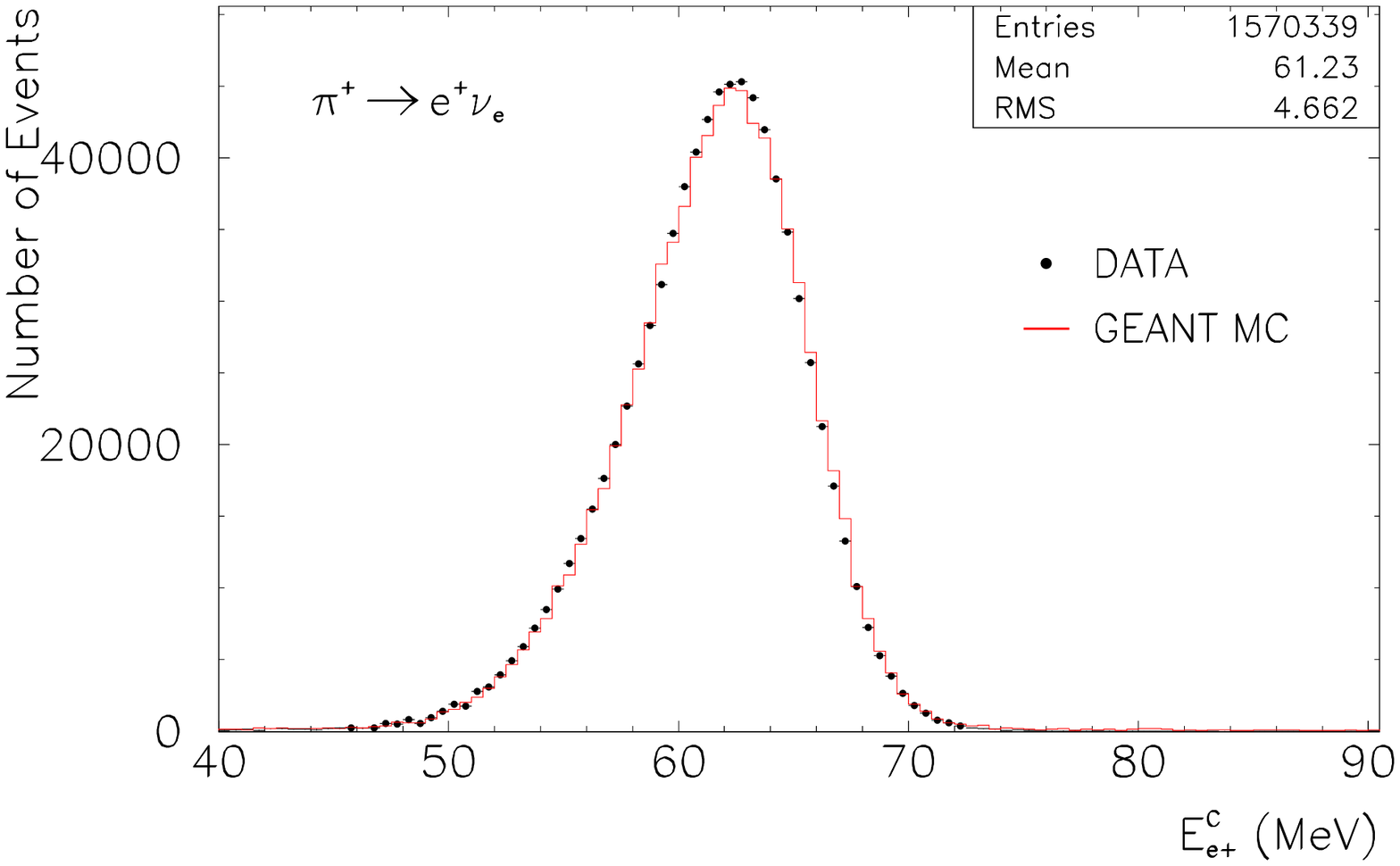}\hspace{0.75cm}
\includegraphics[scale=0.35,origin=c]{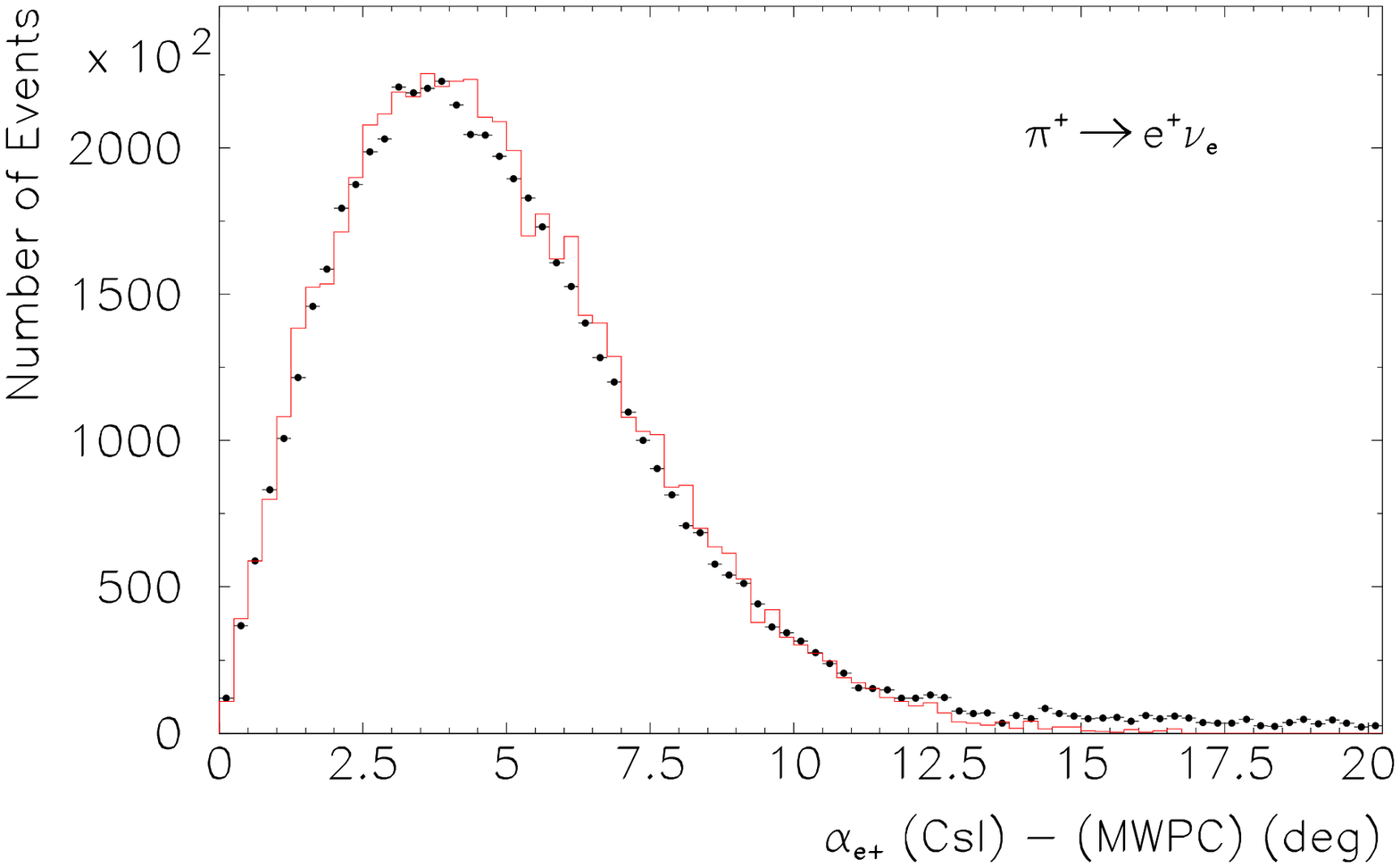}
\end{center}
\caption{A background-subtracted $\pi^+\to e^+\nu_e$ energy spectrum (left).
A track definition using a difference in positron direction measured
with MWPCs and the CsI calorimeter (right).}
\label{fig:p2e_s}
\end{figure}

We have designed fast analog hardware triggers
optimized to accept nearly all non-prompt processes 
contained in the calorimeter with an individual shower energy exceeding the
Michel endpoint (high threshold $\simeq$~52\,MeV), while
keeping the accidental rate to an acceptable level. We have also implemented 
an analogous set of electronically prescaled triggers with
the low threshold of $\simeq$~5\,MeV.
We have run with multiple simultaneous physics and calibration triggers 
at the $\pi^+$ stopping rate of $\sim 8\cdot 10^5\,\pi^+$/s 
as well at a set of reduced beam fluxes down to $4\cdot 10^4\,\pi^+$/s, 
which was crucial for a reliable understanding of the detector response.

An experimental branching ratio $R_i^{\rm exp}$ for a particular pion (muon) decay can be evaluated 
using the expression:
\begin{equation}
\label{eq:br}
{R}_i^{\rm exp}={
{N_i p_i} \over { N_{\pi/\mu} g_{\pi} A_i 
\tau_{l} \epsilon_{\rm PV} \epsilon_{\rm C_1}
\epsilon_{\rm C_2}        
                                     }                    },
\end{equation}
where $N_i$ is the number of the detected events for the process $i$, 
$p_i$ is the corresponding hardware/software prescaling factor (if any), $N_{\pi/\mu}$ 
is the number of the decaying $\pi^+$'s (or $\mu^+$'s), $g_{\pi/\mu}=\int_{t_1}^{t_2}
\exp({-t/\tau})dt$ is the $\pi^+$ ($\mu^+$) gate fraction, $A_i$ is the detector 
acceptance incorporating the specific software cuts, $\tau_l$ is the detector 
live time, $\epsilon_{\rm PV}$ is the plastic veto efficiency, $\epsilon_{\rm C_1}$ 
is the MWPC$_1$ chamber efficiency, and $\epsilon_{\rm C_2}$ is the MWPC$_2$ 
chamber efficiency. In our analysis we have used $t_1=10\,$ns and $t_2=130\,$ns for
the beginning and the end of integration range. The number of decaying $\pi^+$'s is
equal to the number of $\pi^+$'s stopping in the target, corrected for a small 
loss due to hadronic interactions.

\begin{figure} [b] 
\begin{center}
\includegraphics[scale=0.35]{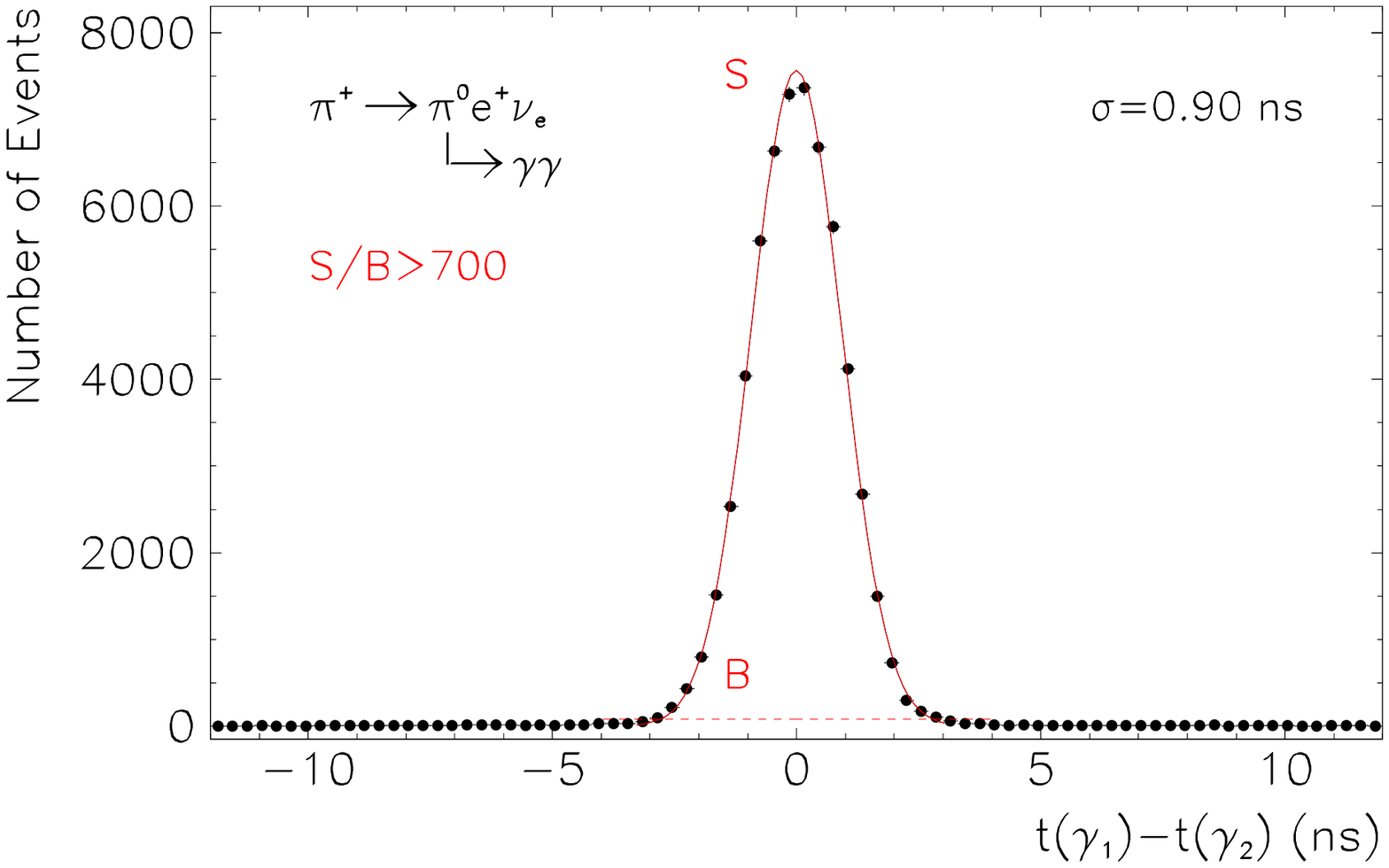}\hspace{0.75cm}
\includegraphics[scale=0.35]{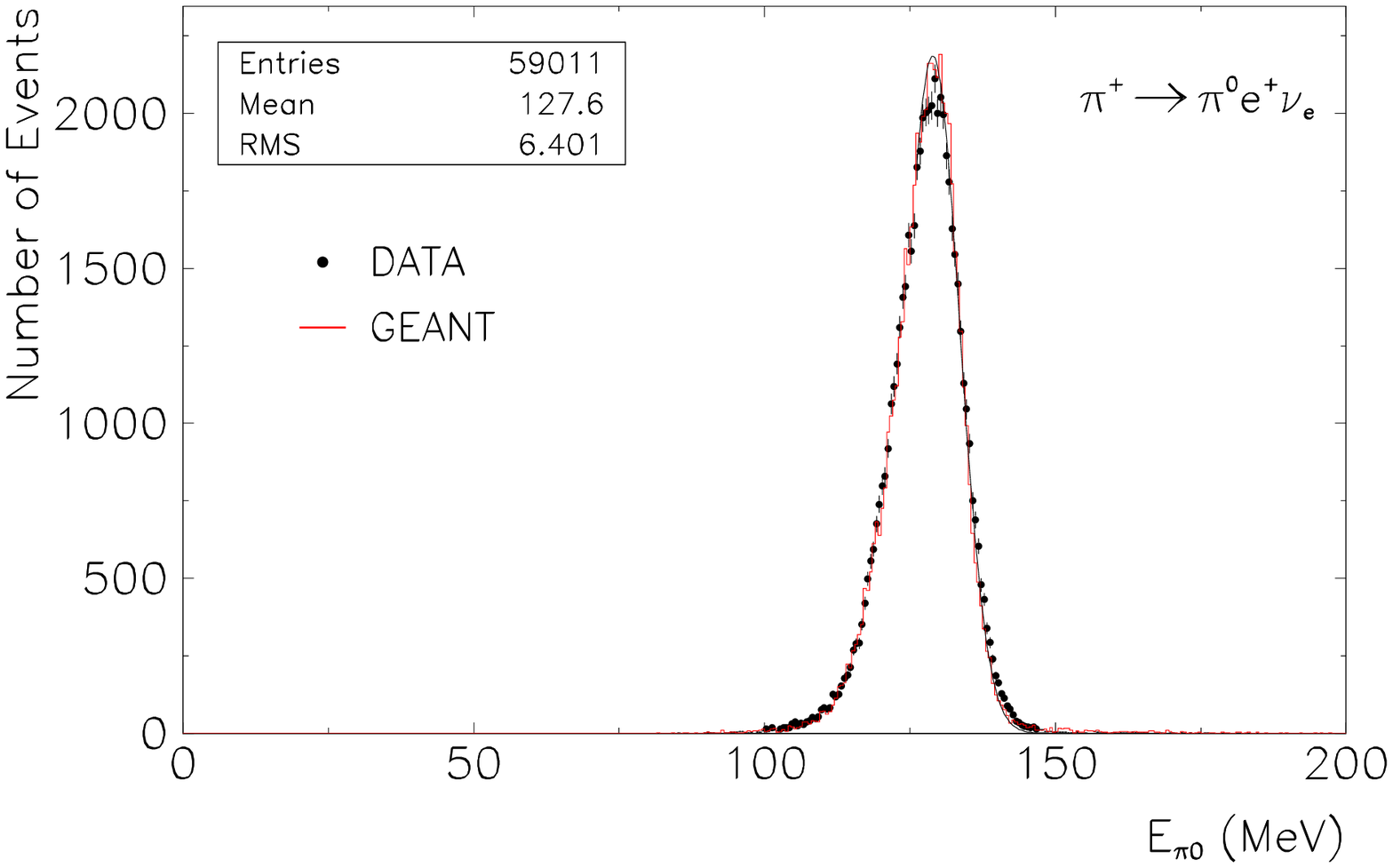}
\end{center}
\caption{The measured timing distribution between two coincident photons
from $\pi\beta$ decays (left). The reconstructed $\pi^0$ energy spectrum 
compared with the Monte Carlo simulation (right).}
\label{fig:pb_s}
\end{figure}

The total yield of the $\pi^+\to e^+\nu_e$ events was evaluated by two independent
methods: (i) from the positron energy spectrum with the Michel background subtracted
using the late-time events, and (ii) by fitting the positron timing spectrum.
The consistency of these two methods was better than $\sim 0.3\,$\%.
The $\pi^+\to e^+\nu_e$ positron energy lineshape and the charged particle
tracking in the wire chambers and the CsI calorimeter are demonstrated in 
Fig.~\ref{fig:p2e_s}.
Using the Eq.~\ref{eq:br} and normalizing to the number of decaying $\pi^+$'s
we find that the $\pi^+\to e^+\nu_e$ branching ratio is independent of the beam
intensity. The average measured $R_{\pi\to e\nu}$ value is:
\begin{equation}
R_{\pi\to e\nu}^{\rm exp}=[1.229\pm 0.003{\rm (stat)}\pm 0.007{\rm (sys)}]\cdot 10^{-4},
\end{equation}
in very good agreement with the theoretical predictions that incorporate
radiative corrections~\cite{Mar93}.

\begin{figure} [t] 
\begin{center}
\includegraphics[scale=0.3]{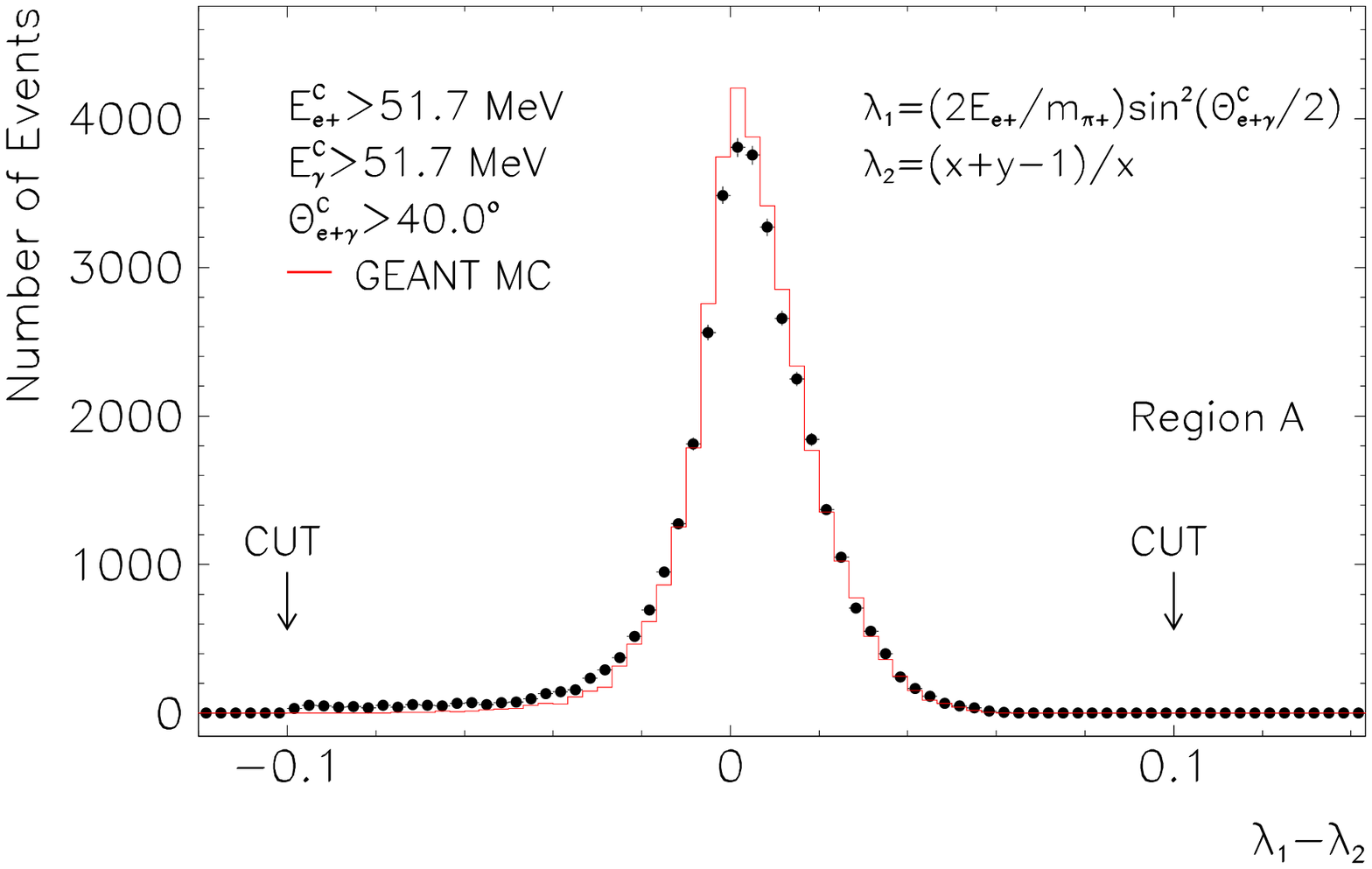}
\includegraphics[scale=0.3]{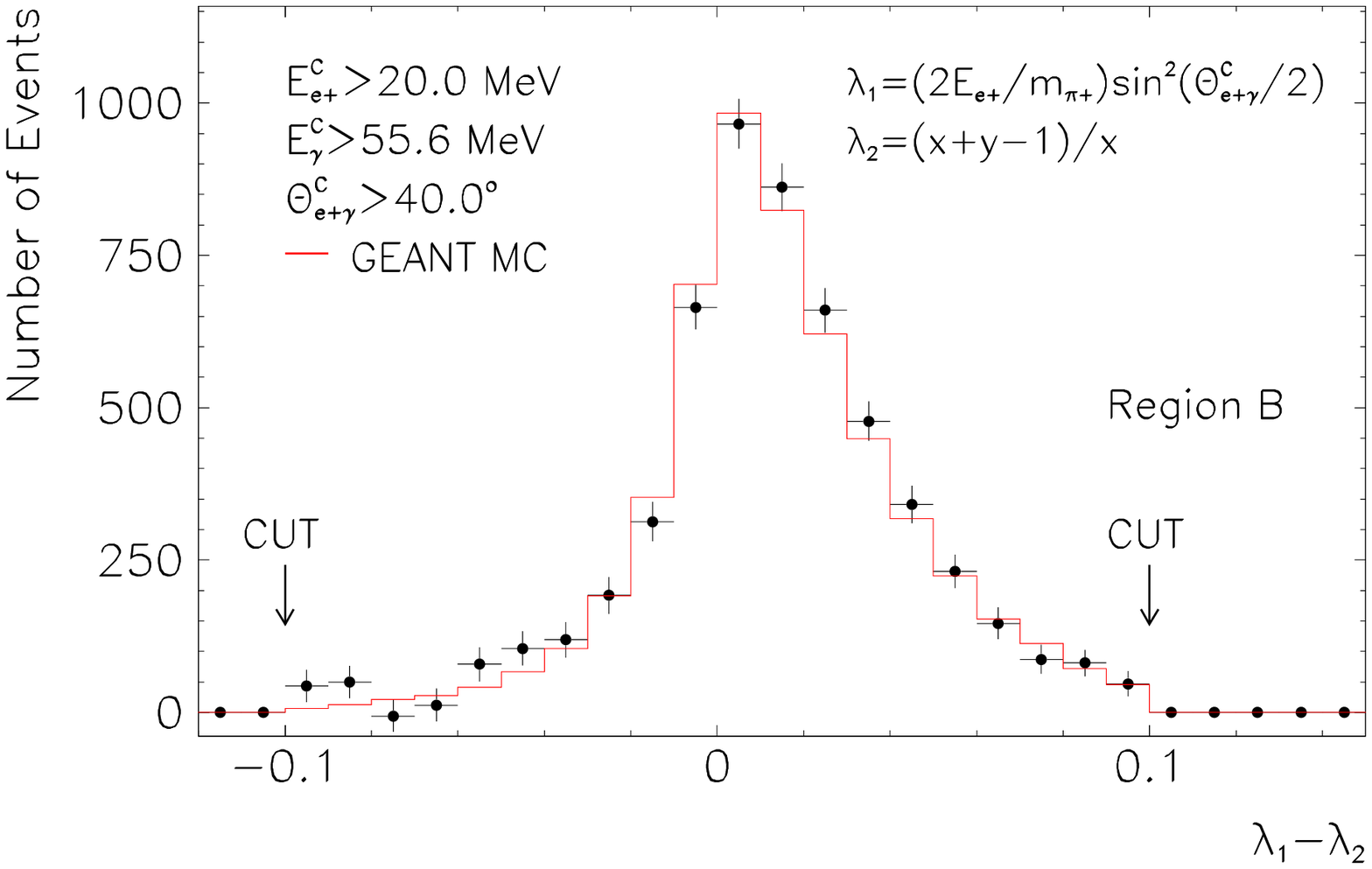}
\includegraphics[scale=0.3]{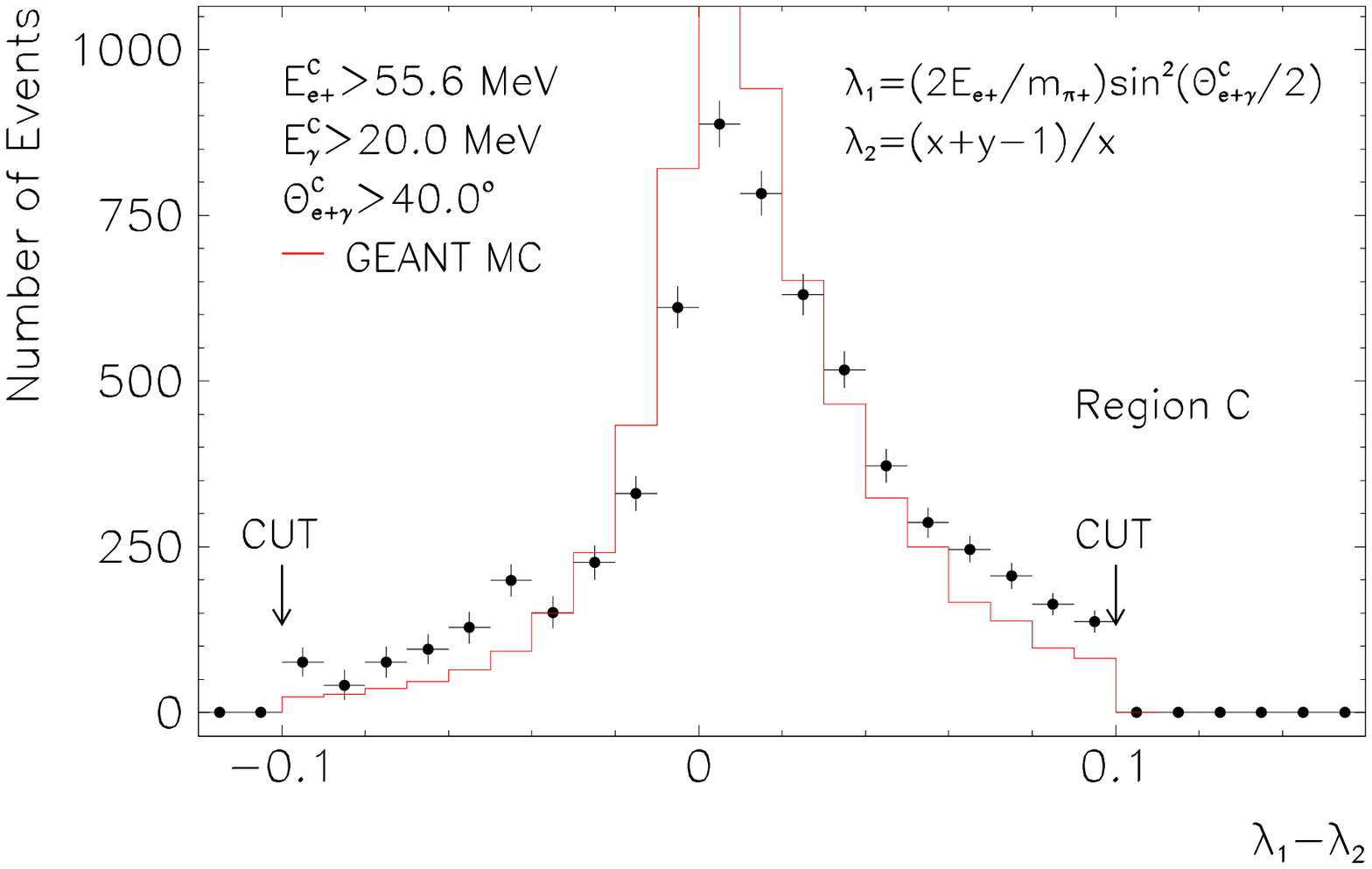}
\end{center}
\caption{$\pi^+\to e^+\nu_e\gamma$: mismatch between the kinematic variable 
$\lambda$ calculated in two alternative ways: (1) from measured positron and 
photon energies $E_{e^+}^{\rm cal}$ and $E_\gamma^{\rm cal}$ and (2) from measured positron 
energy $E_{e^+}^{\rm cal}$ and the opening angle $\theta_{e^+\gamma}^{\rm cal}$ 
(full markers). A Monte Carlo of predicted differences $\lambda_1-\lambda_2$ is shown as 
a full line histogram.}
\label{fig:rp_s}
\end{figure}

As demonstrated in Fig.~\ref{fig:pb_s}, the $\pi\beta$ data sample is exceedingly
pure: the signal-to-background ratio is greater than 700. 
The analysis of the complete statistics in conjunction with
the most stringent off-line cuts yielded $\sim\,$60,000 $\pi\beta$
events. The preliminary branching ratio normalized to the number of
decaying $\pi^+$'s is:
\begin{equation}
R_{\pi\beta}^{\rm exp}=[1.042\pm 0.004{\rm (stat)}
\pm  0.007{\rm (sys)}]\cdot 10^{-8}.
\end{equation}
A consistent $R_{\pi\beta}$ value is obtained when using the known rate 
of the $\pi^+\to e^+\nu_e$ decays~\cite{Mar93} for the absolute normalization.
This method has the lower systematic uncertainties (ultimately $\simeq 0.3\,$\%). 
We note that our experiment tests for the first time the calculation of the 
$\pi\beta$ radiative corrections which stand at ${\rm RC}_{\pi\beta}\sim 
(+3.3\pm 0.1)$\,\%~\cite{rc}.

\begin{figure} [h] 
\begin{center}
\includegraphics[scale=0.35]{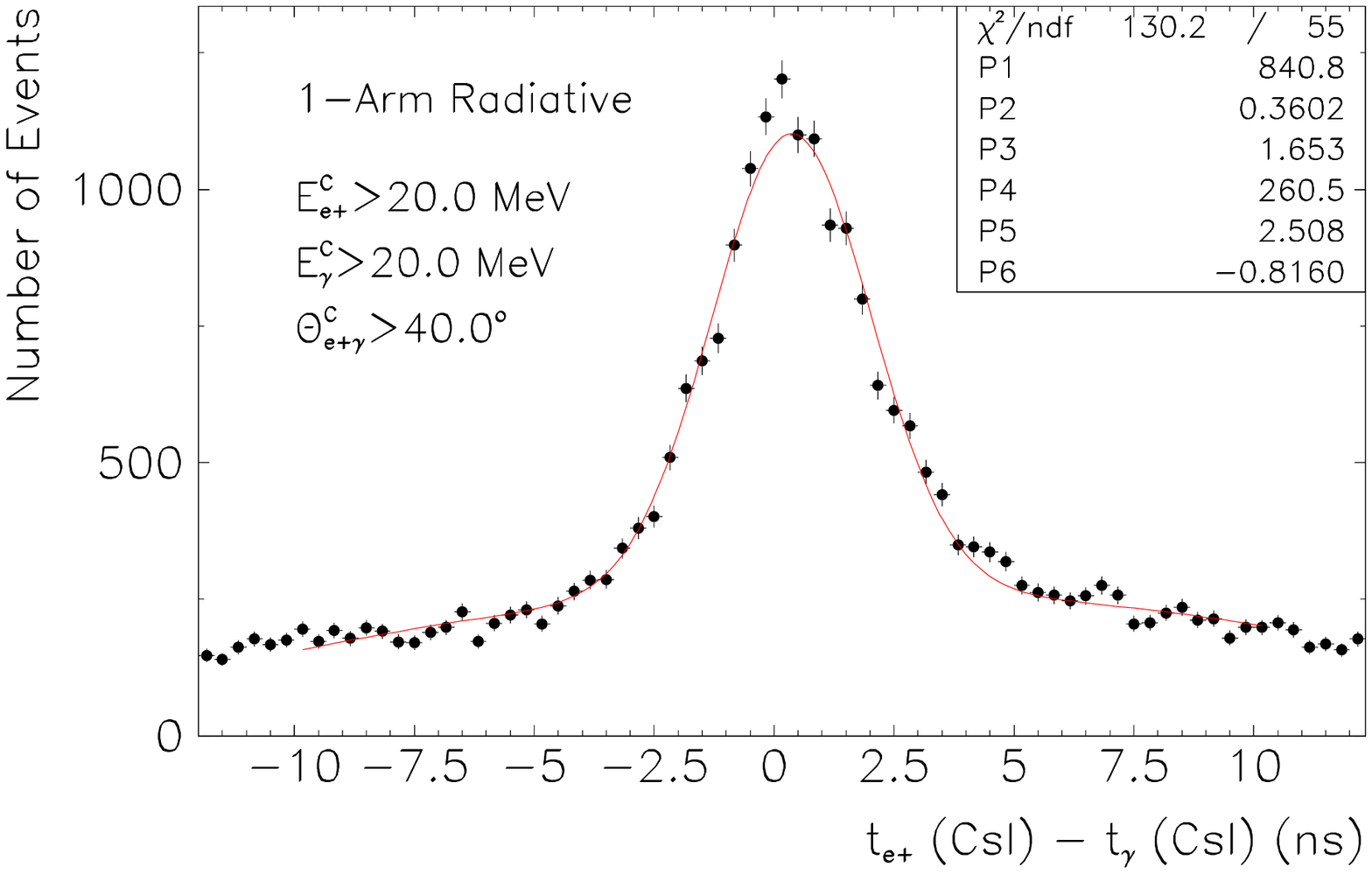}\hspace{0.75cm}
\includegraphics[scale=0.35]{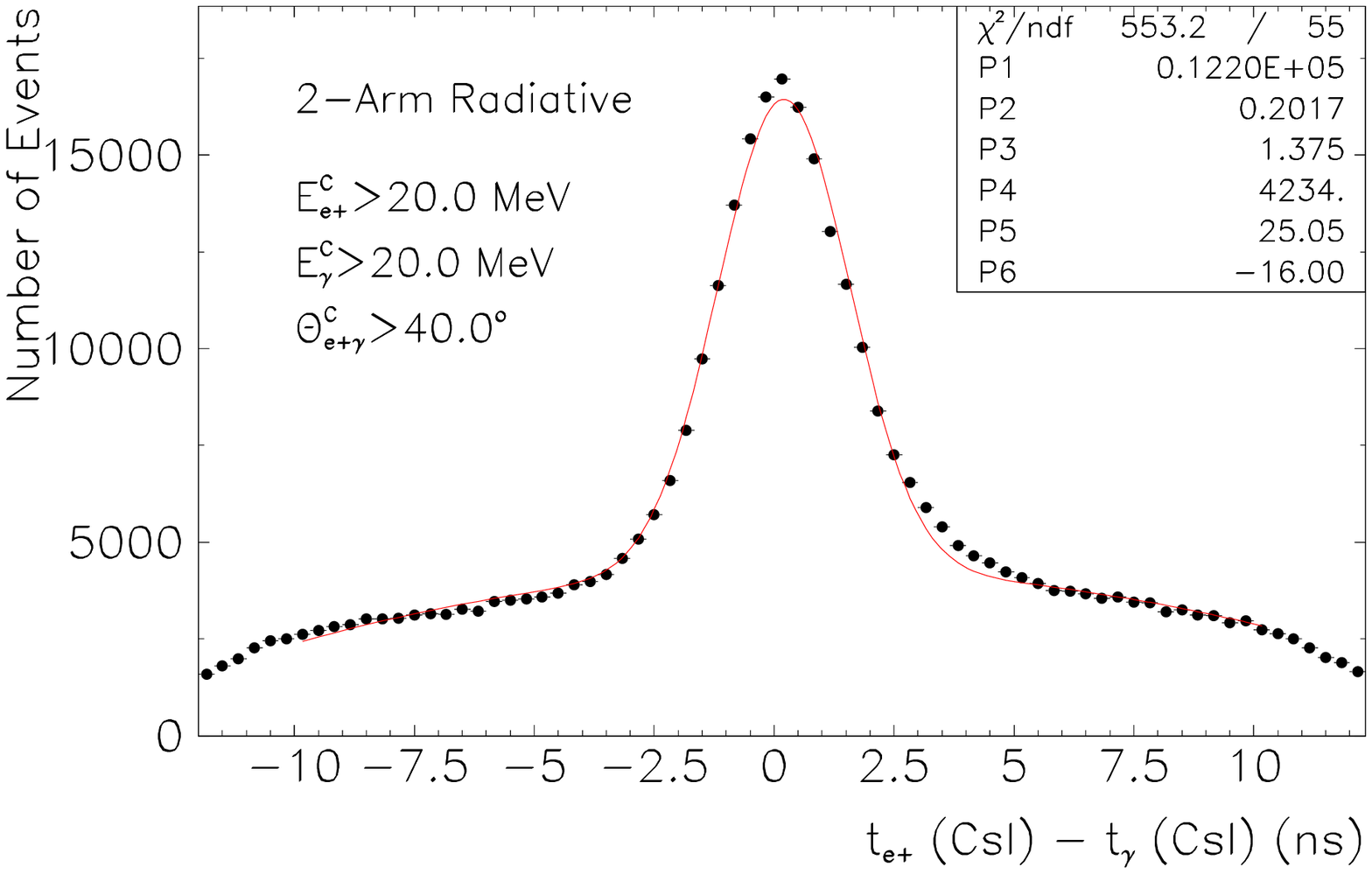}
\end{center}
\caption{Signal-to-background (S/B) ratios for radiative muon events
$\mu^+\to e^+\nu_e\bar{\nu}_\mu\gamma$. The S/B ratio for 1-arm calorimeter  
trigger (left) is 7.5. The 2-arm data (right) have the S/B ratio of 6.0.}
\label{fig:rm_s}
\end{figure}

In its current phase the PIBETA experiment has increased the existing world data set 
for $\pi^+\to e^+\nu_e\gamma$ (RPD) process by more than 30-fold. Using one-arm and 
two-arm calorimeter triggers with high energy threshold we have covered 
the radiative phase space regions dominated by the internal bremsstrahlung process 
as well as by the structure-dependent terms.
The two-arm data set was restricted to 
$e^+$-$\gamma$ coincident pairs for which both measured energies in 
the calorimeter were $E^{\rm cal}_{e^+,\gamma} > 51.7\,$MeV, 
and for which the opening angle $\theta^{\rm cal}_{e^+\gamma} >40.0^\circ$ 
(phase space region $A$). The two one-arm data sets included coincidences 
for which the measured positron (photon) calorimeter energy was 
$E^{\rm cal}_{e^+(\gamma)}>20.0\,$MeV, the photon (positron) energy 
$E^{\rm cal}_{\gamma(e^+)} > 56.4\,$MeV and their opening 
angle $\theta^{\rm cal}_{e^+\gamma} >40.0^\circ$ (phase space regions 
$B$ and $C$).
The reaction yields are calculated by subtracting out-of-time random coincidences
from the events in the $\pm 5\,$ns signal region. The proper accounting
was done for the unavoidable $\pi\beta$ background.
The purity of the final data set is demonstrated in Fig.~\ref{fig:rp_s}.

The corresponding partial branching ratios extracted with
the Monte Carlo minimization algorithm are listed in Table~\ref{tab1}.
The fit corresponds to the ratio of weak axial vector to polar vector form 
factor $\gamma\equiv F_A/F_V$ of
\begin{equation}
\label{eq2}
\gamma_{\rm EXP}=0.443\pm 0.008{\rm (stat)} 
\pm 0.012{\rm (sys)},
\end{equation}
consistent with the present chiral symmetry 
phenomenology~\cite{Hol86}.

Phase space regions $A$ and $C$ agree well with the $(V-A)$ model predictions
and the CVC hypothesis; the region $B$ indicates a $\sim 19\,$\% deficit 
in the number of observed $\pi^+\to e^+\nu_e\gamma$ events.

\begin{figure} [h] 
\begin{center}
\includegraphics[scale=0.35]{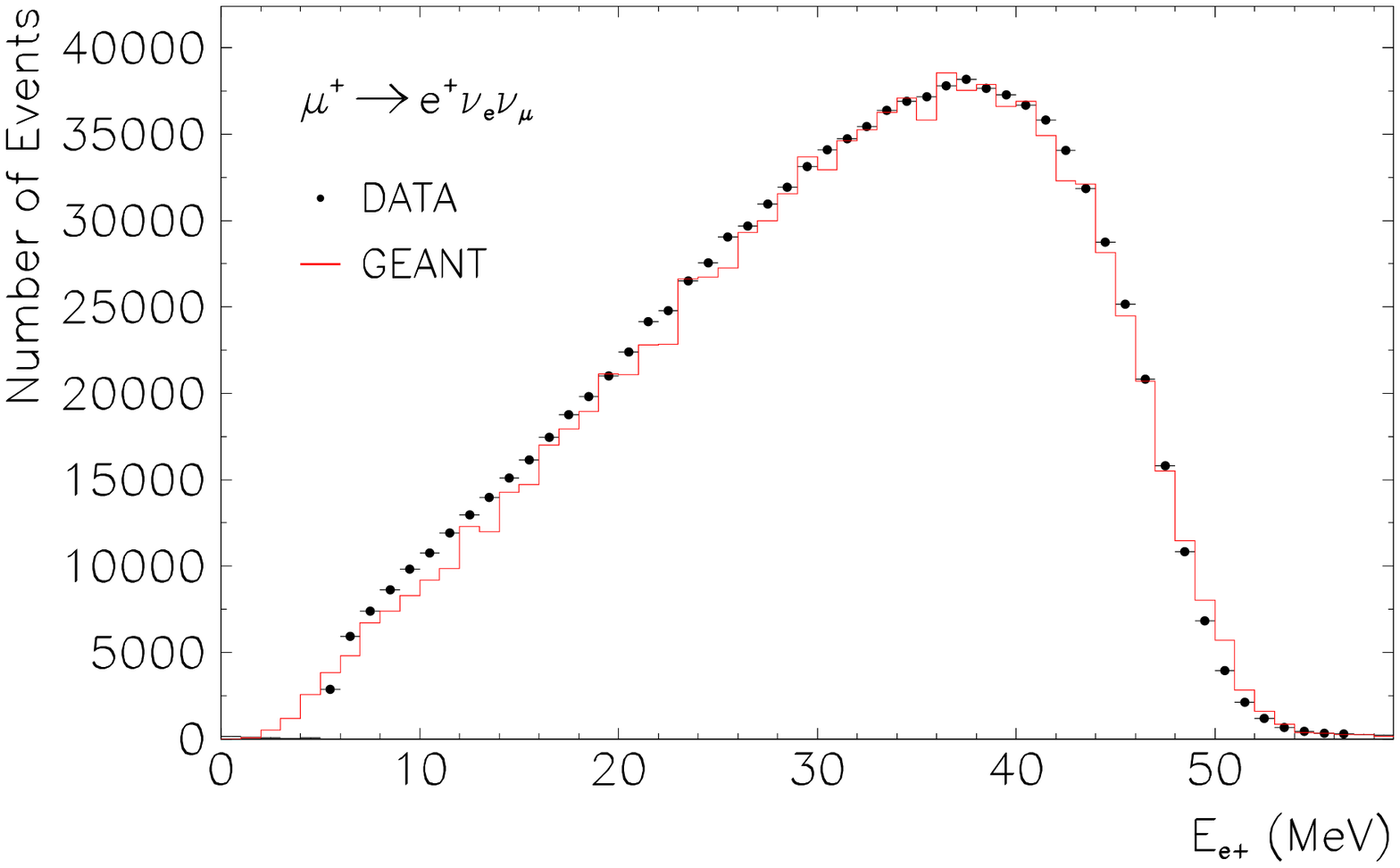}\hspace{0.75cm}
\includegraphics[scale=0.35]{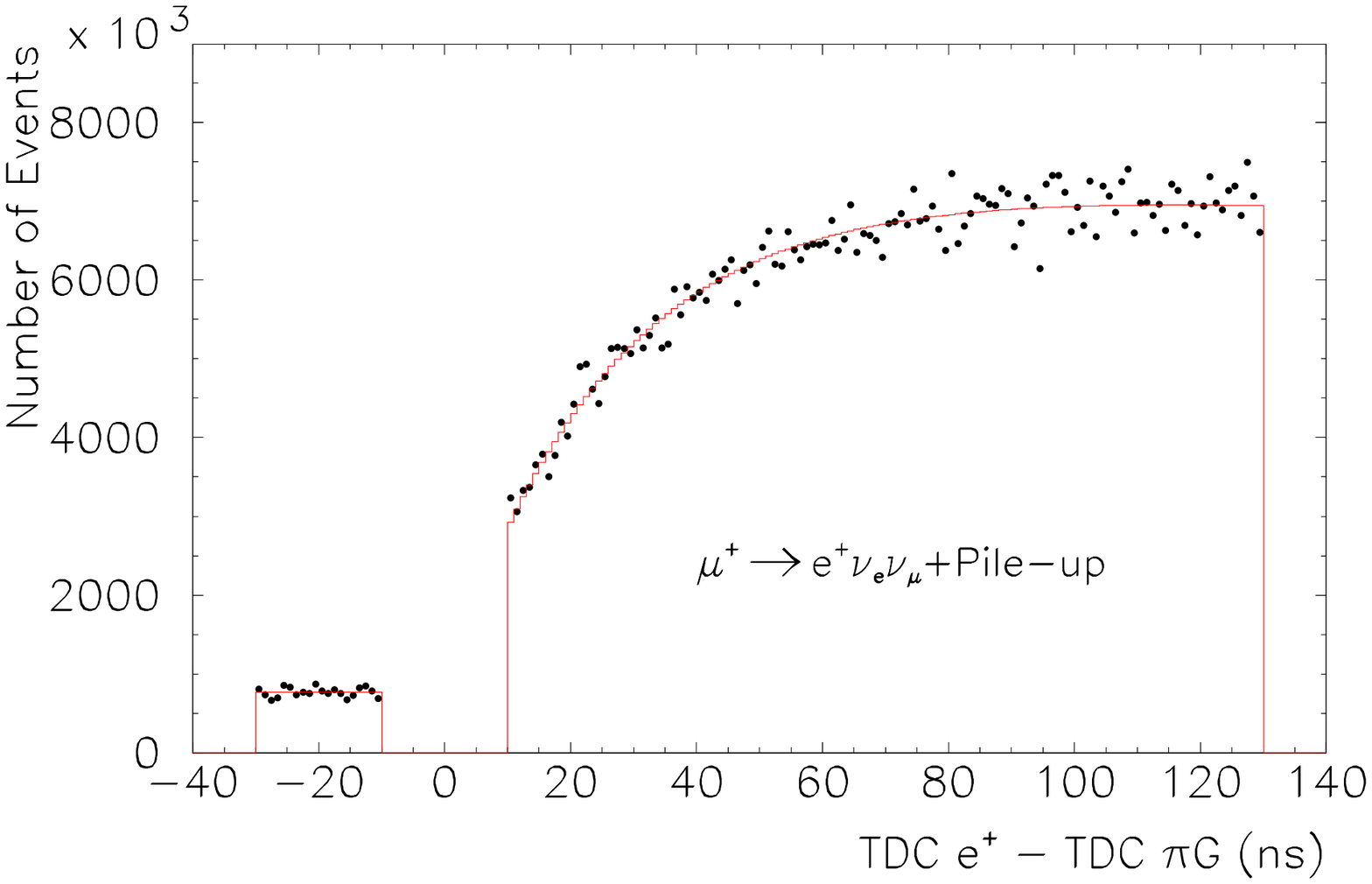}
\end{center}
\caption{The agreement between the measured $\mu^+\to e^+\nu_e\bar{\nu}_\mu$ 
energy spectrum in the CsI calorimeter and the Monte Carlo simulation (left).
The timing spectrum of $e^+$'s from $\pi^+$-$\mu^+$-$e^+$ decay chain at
the beam intensity of $5\cdot 10^4\,\pi^+$/s (right).}
\label{fig:mich_s}
\end{figure}

We have simultaneously recorded a large set of radiative muon decay events $\mu^+\to
e^+\nu_e\bar{\nu}_\mu\gamma$ (RMD) using the prescaled low threshold triggers, Fig.~\ref{fig:rm_s}. 
The experimental branching ratios in Table~\ref{tab1} are calculated from the event yields
and numbers of decaying $\mu^+$'s in the conjunction with the Standard Model 
description~\cite{Eic84} of the process and the Monte Carlo simulation of the detector
response. For the phase space region limited by the positron and photon energies 
$E^{\rm cal}_{e^+,\gamma}>\,20$\,MeV and the particles' opening angle 
$\theta^{\rm cal}_{e^+\gamma}>20^\circ$ the measured radiative muon branching ratio is:
\begin{equation}
R_{\mu\to e\nu\nu\gamma}^{\rm exp}=[2.57\pm 0.05{\rm (stat)}
\pm  0.05{\rm (sys)}]\cdot 10^{-3},
\end{equation}
agreeing with the prediction of the Standard Model~\cite{Eic84}. 
Consistent results are again obtained when
normalization is done with respect to the total number of detected Michel
decays $\mu^+\to e^+\nu_e\bar{\nu}_\mu$, Fig.~\ref{fig:mich_s}. Moreover,
the measured Michel decay branching ratio calculated using Eq.~\ref{eq:br} 
is, within the experimental uncertainties, 100\,\% (Table~\ref{tab1}), indicating the excellent Monte 
Carlo simulation of the detector response as well as properly understood detector 
efficiencies.  

\begin{table}[h]
\caption{Comparison of preliminary experimental and SM-predicted
branching ratios $R_i$. The absolute normalization is done using
the number of decaying $\pi^+$'s or $\mu^+$'s.
In order to allow for a comparison with theoretical $R_i^{\rm the}$'s,
we have imposed further cuts on the positron and photon physical energies for
radiative decay events, in addition to the cuts imposed on the
measured quantities (see text). For RPD: $E_{e^+,\gamma}>50\,$MeV (region $A$),  
$E_{e^+}>10\,$MeV and $E_\gamma>50\,$MeV ($B$),
$E_{e^+}>50\,$MeV and $E_\gamma>10\,$MeV ($C$).
For RMD: $E_{e^+,\gamma}>10\,$MeV.}
\label{tab1}
\begin{center}
\begin{tabular}{lllc}
\hline\hline  \\[-2ex]
Decay     & PIBETA $R_i^{\rm exp}$ Value  & SM Theoretical $R_i^{\rm the}$ & Reference\\[0.5ex]
\hline   \\[-2ex]
$\pi^+\to e^+\nu_e$          & $\rm (1.229\pm 0.003\pm 0.007)\cdot 10^{-4}$ 
                       & $\rm (1.2352\pm 0.0005)\cdot 10^{-4}$ & \protect{\cite{Mar93}} \\
$\pi^+\to e^+\nu_e\gamma$ ($A$) & $\rm (2.71\pm 0.01\pm 0.05)\cdot 10^{-8}$
                       & $\rm (2.58\pm 0.01)\cdot 10^{-8}$     & \protect{\cite{Bry82}} \\ 
$\pi^+\to e^+\nu_e\gamma$ ($B$) & $\rm (1.16\pm 0.02\pm 0.03)\cdot 10^{-7}$
                       & $\rm (1.43\pm 0.01)\cdot 10^{-7}$     & \protect{\cite{Bry82}} \\ 
$\pi^+\to e^+\nu_e\gamma$ ($C$) & $\rm (3.91\pm 0.06\pm 0.12)\cdot 10^{-7}$
                       & $\rm (3.78\pm 0.01)\cdot 10^{-7}$     & \protect{\cite{Bry82}} \\ 
$\pi^+\to \pi^0 e^+\nu_e$    & $\rm (1.042\pm 0.007\pm 0.009)\cdot 10^{-8}$ 
                        & $(1.039\pm 0.001)\cdot 10^{-8}$      & \protect{\cite{rc}} \\ 
$\mu^+\to e^+\nu_e\bar{\nu}_\mu$ & $\rm 0.971\pm 0.003\pm 0.010$
                        & $0.988\pm 0.005$                     & \protect{\cite{Hag02}} \\
$\mu^+\to e^+\nu_e\bar{\nu}_\mu\gamma$ 
                  & $\rm (2.57\pm 0.05\pm 0.05)\cdot 10^{-3}
                         $ & $(2.584\pm 0.001)\cdot 10^{-3}$   & \protect{\cite{Eic84}} \\[1ex]
\hline\hline
\end{tabular}
\end{center}
\end{table}
\section{\sl Acknowledgements}
The PIBETA experiment has been supported by the National Science Foundation,
the Paul Scherrer and the Russian Foundation for Basic Research.
This material is based upon work supported by the National Science
Foundation under Grant No.\ 0098758.

\end{document}